\documentstyle[12pt,a4,epsf]{article}
\begin{document}

\baselineskip15pt

\thispagestyle{empty}

\begin{flushright}
\begin{tabular}{l}
FFUOV-97/01\\
{\tt hep-th/9701122}
\end{tabular}
\end{flushright}

\vspace*{2cm}

{\vbox{\centerline{{\Large{\bf THE STRING DENSITY OF STATES 
}}}}}

\vspace{0.8cm}

{\vbox{\centerline{{\Large{\bf 
FROM THE CONVOLUTION THEOREM}}}}}

\vskip30pt

\centerline{Marco Laucelli Meana, M.A.R. Osorio, and Jes\'{u}s Puente Pe\~{n}alba 
\footnote{E-mail address:
     laucelli, osorio, jesus@string1.ciencias.uniovi.es}}

\vskip6pt
\centerline{{\it Depto. de F\'{\i}sica, Universidad de Oviedo}}
\centerline{{\it Avda. Calvo Sotelo 18}}
\centerline{{\it E-33007 Oviedo, Asturias, Spain}}

\vskip .25in

\vspace{1cm}
{\vbox{\centerline{{\bf ABSTRACT }}}}

We study the microcanonical density of states and the thermal 
properties of a bosonic string gas starting from a calculation
of the Helmholtz free energy in the S-representation. By adding more and more strings
to the single string system, we induce that, for infinite volume,
there is no negative specific heat region but a transition at a finite value of the energy per
string  from the low energy regime 
to a region of infinite
specific heat at the Hagedorn temperature. Forcing the description of this phase in terms of
strings gives a picture in which there is a very fat string in a sea of low energetic ones.
We argue that the necessary changing of this description should not change the fact that
perturbatively $T_H$ is a maximum temperature of the system.

\vspace*{24pt}

\setcounter{page}{0}
\setcounter{footnote}{0}

\newpage

\section{Introduction}
What could be called the Hagedorn problem in String Theory has been a long-standing
issue that lasts  for more than twenty five years \cite{FrauCarCabbiHage}.
Many of us at some
moment in the recent past decided to, more or less, forget about it based on the
belief that non perturbative string theory should have something to say on this
question. Now that we are able  to compute D-instanton effects \cite{BarboVazquez}
\cite{Green}
it seems very clear
that the influence of them on the microcanonical density of states could  be of some
interest. The main question is to know whether there is some sensible modification of
the picture obtained from the perturbative regime.

First of all, it would be mandatory to collect all the references about different
computations of the microcanonical density. There are many of them
(cf. for example \cite{DeoAxeMac}\cite{Alv}). We focused our
interest on the computations we consider as more fundamental in the sense that
 to get the density of states one makes the inverse Laplace transform of
 the canonical partition function. This is actually the definition of the
microcanonical ensemble when using quantum statistics \cite{Horo}.
 Problems arise soon. The
first one is that of choosing a  one loop calculation of 
$Z(\beta)=e^{-V\beta {\cal F}(\beta)}$ by taking either what is called the
F-representation or the S-representation \cite{MacRothOsoKutaSeiberg}. 
Both are in fact different from one another
 and only coincide in a part of the complex plane 
 once they are analytically continued as functions of the temperature. To be
more precise, the F-representation does not have a unique continuation
\cite{vaz1vaz2} to the complex
plane and so cannot be used to make the inverse Laplace transform. In general, in the
F-representation, in which modular invariance is manifest,
there are many singularities away from the real and imaginary axes that do not appear 
in the S-representation. We have then to use the S-representation in which  the Helmholtz 
free energy  is computed as the sum over the field content of the string.

Up to our present knowledge, all the calculations in the literature have been carried
out taking from the very beginning a mostly vague high energy limit to obtain an asymptotic form
for the  density of states which does not contain quantum statistic ingredients.
Now the question we pose is a little bit naive, but to us, will show up as the most
important aspect of the problem: how can we connect the low energy regime with this
 high energy behaviour? It seems to be clear that a string has two different types of
excitations, the massless ones which configure what is called the effective theory
and all the modes with masses multiple of the Planck mass. The massless modes should
dominate the low energy region and the massive excitations, once the threshold is
opened, should be responsible for any bizarre behaviour in the high energy part.

We started thinking about computing D-instanton effects on the density of states, but
after reviewing the contribution from the perturbative series,
we have found many things we consider are important enough so as to  postpone
the original plan. Nevertheless a sketch about the influence of non
perturbative effects will be depicted in the  conclusions. In this paper  we
 shall be devoted to a revision of the computation of the string density of states for
the closed bosonic string using the S-representation. First, we compute the
density of states for a single string and by repeatedly using the convolution theorem
 we get the density of states for $N$ strings.

\section{Free energy of the bosonic string  in 
\protect\newline the S-representation}

 The free energy  per unit volume for a bosonic string gas in the  S-representation,
also called analog model,  is 
\cite{mar}
\begin{equation}
\hspace{0mm}
{\cal F}(\beta)=-\frac{2^{-14}\pi^{-26}}{(2\alpha')^{13}}\int_{0}^{\infty}d
\tau_2\tau_2^{-14}\left[\theta_3\left(0,\frac{i\beta^2}{4\pi^2\tau_2e
\alpha'}\right)-1\right]\int_{-1/2}^{+1/2}d\tau_1\left|\eta(\tau)\right|^{-48}
\label{fbeta}
\end{equation}
where $\tau=\tau_1+i \tau_2$. The integral in $\tau_1$ may be written as:
\begin{equation}
B(\tau_2)=\int_{-1/2}^{+1/2}d\tau_1\left|\eta(\tau)\right|^{-48}=\sum_{k=-1}^{\infty}a_k^2e^{-
4\pi k\tau_2}
\end{equation}

There, we have used the following expansion for the $-24^{th}$ power 
of the  Dedekind eta function

\begin{equation}
\eta\left({\tau}\right)^{-24}=\sum_{n=-1}^{\infty}a_n e^{2\pi i n\tau}.
\label{etaseries}
\end{equation}

Writing the Jacobi function $\theta_3(0,\tau)$ as a series
\begin{equation}
{\cal F}(\beta)=-\frac{2^{-14}\pi^{-26}}{(2\alpha')^{13}}
\sum_{r=1}^{\infty}\sum_{k=-1}^
{\infty}a_k^2\int_{0}^{\infty}d\tau_2\tau_2^{-14}e^{-\frac{\beta^2 r^2}{4\pi\tau_2
\alpha'}}e^{-4\pi k\tau_2}
\end{equation}
and using an integral expression of the $K_{d/2}$ modified Bessel function,
 we have  
\begin{equation}
\hspace{-0.5mm}-\beta {\cal F}(\beta)=\frac{2^{-12}\pi^{-26}}{(2\alpha')^{-13}}
\sum_{r=1}^{\infty}
\sum_{k=-1}^{\infty}a_k^2(4\pi k)^{13}\beta\left(\sqrt{\frac{k}{\alpha'}}\beta r\right)
^{-13}K_{13}\left(2\sqrt{\frac{k}{\alpha'}}\beta r\right)
\label{K13}
\end{equation}

Where the $k=-1$ term comes from the tachyon and will be neglected because it
produces an infrared divergence; i.e., the tachyon shows its negative squared mass 
when the proper time goes to infinity.
%
%
The partition function can be expanded as
\begin{equation}
Z(\beta)-1=\sum_{n=1}^{\infty}\frac{\left(-V\beta {\cal F}(\beta)\right)^n}{n!}.
\label{Zexpand}
\end{equation}
On the other hand, the density of states as a function of the volume and the total
energy of the system is defined as 
\begin{equation}
\Omega(E)=\frac{1}{2 \pi i}\int_Cd\beta e^{\beta E} Z(\beta)
\end{equation}
where $C$ is a Bromwich contour, so the density of states is calculated as the inverse 
Laplace transform of the partition function. If we assume
 that our system is made of many subsystems,
that we will interpret afterwards as strings, the density of states 
can also be written as \cite{vaf}

\begin{equation}
\Omega(E)=\sum_{n=1}^{\infty}\frac{1}{n!}\int_0^{\infty}\prod_{i=1}^{n}dE_i
\Omega_1(E_i)\delta\left(\sum_{i=1}^{n}E_i-E\right)
\end{equation}
where $\Omega_1(E_i)$ is the density of states of the single string subsystem.
We write $\delta(x)$ as a Fourier integral and we obtain

\begin{equation}
\Omega(E)=\sum_{n=1}^{\infty}\frac{1}{n!}\left[\frac{1}{2 \pi i}\int_C d\beta
e^{\beta E}z(\beta)
^n\right]
\label{omegaexpand}
\end{equation}

Taking the inverse Laplace transform of (\ref{Zexpand}), 
it is easy to see that \linebreak
${\cal L}^{-1}\left\{-V \beta {\cal F}(\beta)\right\}$ is the
 density of states of the single string.
This way, the inverse of the $n^{th}$ power would be 
the microcanonical density of the gas with $n$ strings.
This allows us to relate term by term equations (\ref{Zexpand}) and
(\ref{omegaexpand}), to give
\begin{equation}
\Omega(E)=\sum_{n=1}^{\infty}\Omega_n(E)=\sum_{n=1}^{\infty}
\frac{V^n}{n!} {\cal L}^{-1}\left\{\left(-\beta {\cal F}(\beta)\right)^{n} \right\}
\label{omegan}
\end{equation}

It is worth mentioning that the chemical potential of our system is zero expressing
the fact that the number of subsystems (strings) can change and then we have to
perform the summation.


For a certain finite volume, there is always a finite $\tilde{n}$ for which 
the corresponding term in (\ref{omegan}) is leading.
That would be the most probable number of strings. Since in our problem 
$V\rightarrow\infty$, $\tilde{n}$ also goes to 
infinity, because the last term we consider is always the biggest. We will go back
to this point later on.


Usually, when a calculation of the inverse Laplace transform of a product is needed,
the convolution theorem can be used. Mathematically, it says

\begin{equation}
{\cal L}^{-1}\left\{f_1(x)f_2(x)\right\}=\int_{0}^{p}\tilde{f}_1(p-t)\tilde{f}_2(t)dt
\end{equation}
in our case:
\begin{equation}
\Omega_2(E)=\frac{1}{2!}\int_{0}^{E}\Omega_1(E-t)\Omega_1(t)dt
\label{omega2}
\end{equation}
that is equivalent, physically, to summing up over every possible 
distribution of the total energy of the
system beetwen the two subsystems. Remember that for a concrete 
energy distribution the number of
microscopic  states is the product of the ones of each subsystem.

\section{\mbox{\Large{\bf{$\Omega_1(E)$}}} at high energies}  

A calculation of the microcanonical density of states at high energy can be found,
for example in \cite{DeoAxeMac}, obtaining the
 leading term of the asymptotic series taking only into account the ultraviolet 
singularity of the free energy at the Hagedorn temperature.
We have seen that it is possible to obtain the whole series
considering all the quantum effects, that
is, all the Hagedorn type singularities on the real axis of the free energy. 
We write the Helmholtz free energy as in
$(\ref{fbeta})$, and note that the part that leads the behaviour
 of the whole integral at high energy 
is the ultraviolet one, that is:
\begin{equation}
-\beta {\cal F}(\beta)\simeq-\frac{2^{-15}\pi^{-26}}{(2\alpha')^{13}}\int_{0}^{\varepsilon}
d\tau_2\tau_2^{-14}\left[\theta_3\left(0,\frac{i\beta^2}{4\pi^2\tau_2
\alpha'}\right)-1\right]e^{\frac{4\pi}{\tau_2}}\tau_2^{51/2}    
\end{equation}

Where the integral of the Dedekind eta function has been approximated in the following
 way. 
Using the inversion properties of the eta function, it is possible to write

\begin{equation}
B(\tau_2)=\int_{-1/2}^{+1/2}d\tau_1\left|\tau\right|^{-24}\left|\eta\left(-\frac{1}{\tau}\right)
\right|^{-48}.
\end{equation}

The same Fourier expansion of the $-24^{th}$ power we used in
(\ref{etaseries}) yields 

\begin{equation}
B(\tau_2)=\sum_{n,m=-1}^{\infty}a_n a_m\int_{-1/2}^{+1/2}d\tau_1\left|\tau\right|^{-24}
 e^{\frac{2\pi i \tau_1}{\left|\tau\right|^2} (m-n)} e^{-\frac{2\pi\tau_2}{\left|\tau\right|^2} (m+n)}.
\end{equation}

 Now, we approximate the series by its first term, defined by $n=m=-1$, because it is the
 only divergent one as $\tau_2\rightarrow 0$. Expanding the integrand around $\tau_1=0$,
 which is the region which encloses most of the area in the integral:
 
\begin{equation}
B(\tau_2) \simeq \frac{1}{2}e^{\frac{4\pi}{\tau_2}}\tau_2^{51/2} 
\end{equation}
which is a better approximation as $\varepsilon$ approaches $0$. To finally evaluate the
 resulting integral in $\tau_2$ it is 
useful to change $\tau_2$ by $1/\tau_2$, and  we also eliminate the tachyon term.
 Using the
integral representation of the incomplete Gamma function, the following expression 
for the single string
partition function is found
\begin{equation}
\hspace{-1mm}
Z^{(1)}(\beta)=\frac{V\beta}{2(2\pi)^{
26}\alpha'^{13}}\sum_{n=1}^{\infty}\left[\frac{4\pi}{\beta_n^2}
\left(\beta^2-\beta_n^2\right)\right]^{25/2}\Gamma\left(-\frac{25}{2},\frac{4\pi}{\varepsilon\beta_n^2}
\left(\beta^2-\beta_n^2\right)\right)
\end{equation}
where $\beta_n=\frac{4\pi\sqrt{\alpha'}}{n}$. When $n=1$ we have the inverse of
 the Hagedorn temperature that we will call  $\beta_H$ from 
now on. Continuing each term of this series to the complex $\beta$ plane it is easy to see
that it has branch points at every $\beta_n$, but every term of the series is well 
defined with isolated singularities at the branch points and the corresponding branch cuts. 
This allows us to invert Laplace term by term looking at each term as an independent inverse
Laplace transfom problem, what guarantees
the convergence of the resulting series, which is
\begin{equation}
\Omega^{(1)}(E)=\sum_{n=1}^{\infty}\sum_{p=0}^{1}\sum_{m=0}^{\infty}A(m,n,p)\frac{e^{\beta_n
E}}{E^{27/2+m+p}} {\cal H}\left(E-\frac{8\pi n}{\beta_H \varepsilon}\right)
\label{highedensity}
\end{equation}
where 
\begin{eqnarray}
A(m,n,p) & = &
V\,2^{21/2-3m-p}\,\pi^{-27/2}\,\alpha'^{-13}n^{23/2+p+m}\,\beta_H^{-23/2-m-p}\nonumber \\
& \times & (-1)^{m+p}\frac{(27+2p+2m)!!}{m!(25-2m)!!}
\end{eqnarray}
and ${\cal H}(x) $ is the Heaviside function. This is a finite series
in the $n$ index and an asymptotic one in $m$. The leading term is the well known Hagedorn one,
 that accounts for
the Maxwell-Boltzmann statistics. The series does not fundamentally change the behaviour of this
term, but it corrects the apparent divergence that appeared  in previous calculations at
 low energy because the step functions 
eliminate its contribution there.  After all, at high energy Bose and Maxwell-Boltzman statistics
are expected to be equivalent.
Nonetheless this density
 of states of the single string cannot be used to obtain
the multiple string through convolution because it does not have the information of the low energy
range which is completely contained in the infrared  part
neglected in the above calculation by introducing the cut off $\varepsilon$.
It is important to notice that although the behaviour of a function in the complex plane is
nearly completely defined by its singularities, the regular parts do play an important
 r\^{o}le when we Laplace
 transform. As an example, in the above calculation the step functions
 that came from the analytical
parts guaranteed the convergence of the series in the $n$ index.

\section{Complete single string density of states}
In this section we will obtain the complete single string density of states,
 and compare it with 
$\Omega^{(1)}(E)$ obtained in (\ref{highedensity}). We will
 start from equation (\ref{K13}). To perform the inverse Laplace transform
 we use the formulas in \cite{ede} to get
%
\begin{eqnarray}
& {\cal L}^{-1} &\hspace{-3mm}\left\{-\beta
{\cal F}(\beta)\right\}=\frac{2^{-12}\pi^{-26}}{(2\alpha')^{-13}}\sum_{r=1}^{\infty}
\sum_{k=-1}^{\infty}a_k^2(4\pi k)^{13}\left(\sqrt{\frac{k}{\alpha'}} r\right)^{-13}\nonumber 
\\
& \times &\left({\cal L}^{-1}\left\{\beta^{-13}K_{13}
\left(2\sqrt{\frac{k}{\alpha'}}\beta r\right)\right\}
\ast{\cal L}^{-1}\{\beta\}\right)
\end{eqnarray}

Remebering that ${\cal L}^{-1}\{\beta\}$ equals minus the derivative of the
Dirac Delta function we obtain
\begin{equation}
\label{compom}
\Omega_1(E)=\frac{V\pi^{-51/2}}{\Gamma\left(\frac{25}{2}\right)2^{38}}
\sum_{r=1}^{\infty}\sum_{k=0}^{\infty}r^{-26}a_k^2
E\left(E^2-\frac{4 r^2 k}{\alpha'}\right)^{23/2}{\cal H}\left(E^2-\frac{4 r^2 k}{\alpha'}\right)
\vspace{4mm}
\end{equation}

The most innovative thing is that in our calculation it is explicit the relativistic nature of the
theory. The Heaviside step functions represent the abrupt appearing of the mass levels
because while there is not enough energy to create  massive fields, those states are not accesible.
The Hagedorn singularity, explicit in the other calculation,
 is here hidden in the enormous
growth of the $a_k^2$ coefficients, that represent the degeneracy of the mass levels. 
This  expression for the density of states is valid over the whole range of values of the
total energy of the system; for example, it possible with it to study 
the massless quantum field regime of the
theory and the transition to the "Hagedorn phase". In fact, for energies lower than the Planck 
mass the series has only the $k=0$ term
so that the system behaves exactly as a collection of massless quantum fields.
With the density of states in eq.(\ref{compom}) the multiple string microcanonical density 
 can be gotten using convolution theorem so as to numerically compute it. Notwithstanding,
in the case of the effective theory, it is 
easy to obtain a simple expression. The single particle density of states is 
\begin{equation}
\Omega_{1,\,eff}(E)=\frac{V}{2^{38}\pi^{51/2}}\zeta(26)(24)^2 E^{24}
\end{equation}
where $\zeta(z)$ is Riemman's Zeta function. Note that $a_0^2=24^2$ represents the
degeneracy of  the massless level. Twenty four is the number of transverse dimensions.
 If we make $N$ convolutions:
\begin{equation}
\Omega_{N,\,eff}(E)=
\frac{1}{\Gamma(25N)N!}\left(\frac{V}{2^{38}\pi^{51/2}}\zeta(26)(24)^2\Gamma(25)\right)^N E^{25N-1}
\end{equation}
which is the micraconical density for $N$ massless particles in 25  spatial dimensions.
We can use this expression as a model to understand the relation between the volume and the number
of subsystems  --in this case, particles. For a given volume we can calculate, in terms
of $N$, the most degenerate configuration of the gas, which is the most probable. We obtain the most
probable value of the number of subsystems by evaluating $\tilde{N}$ which is the most relevant term
of the series obtained summing up all the possible $N$'s:
\begin{equation}
\tilde{N}\propto u^{25} V
\end{equation}
where  $u$ is the energy per particle. If we take the volume to infinity, the number $\tilde{N}$
 goes also to infinity, but the density of particles per unit volume remains finite as long as
 the energy per particle  is finite.

\section{ The problem of energy equipartition}

To study the most probable distribution of the total energy of the system among 
the strings we will write
the density of states for a given energy distribution. Let~ us imagine a system with two strings
that share a total energy $E$, when one of them has an energy ${\cal E}$,
the microcanonical density is
\begin{equation}
\omega(E,{\cal E})=\frac{1}{2!}\Omega_1(E-{\cal E})\Omega_1({\cal E})
\end{equation}
which is the integrand and combinatorial factor in (\ref{omega2}).
The probability would be
\begin{equation}
{\cal P}(E,{\cal E})=\frac{\frac{1}{2!}\Omega_1(E-{\cal E})\Omega_1({\cal E})}{\Omega_2(E)}
\end{equation}

$\Omega_1(E)$ is the single string density of
states that we have already calculated. Loocking at this distribution in terms of ${\cal E}$
 for a given
$E$ we see that the behaviour is very different whether the total energy is bigger or smaller than
$E_0 =  2 u_0 \simeq \frac{2.8}{\sqrt{\alpha'}}$.

\begin{figure}
\let\picnaturalsize=N
\def\picsize{3in}
\def\picfilename{2cuerda.eps}
\ifx\nopictures Y\else{\ifx\epsfloaded Y\else\input epsf \fi
\let\epsfloaded=Y
\centerline{\ifx\picnaturalsize N\epsfxsize \picsize\fi
\epsfbox{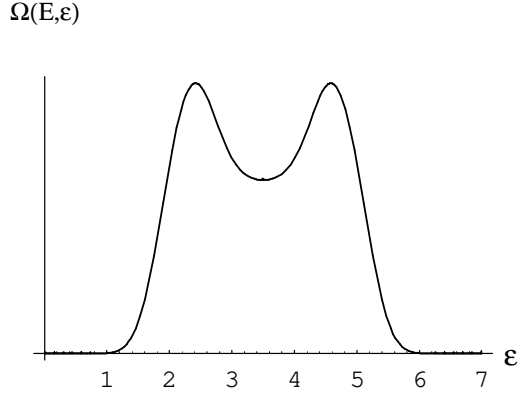}}}\fi
\caption{Degeneracy of the two string system, in the broken-equipartition regime. The total
energy of the system is $\frac{7}{(\alpha^{\prime})^{1/2}}$.}
\label{degeneracy2}
\end{figure}

\begin{figure}
\let\picnaturalsize=N
\def\picsize{2in}
\def\picfilename{3cuerda.eps}
\ifx\nopictures Y\else{\ifx\epsfloaded Y\else\input epsf \fi
\let\epsfloaded=Y
\centerline{\ifx\picnaturalsize N\epsfxsize \picsize\fi
\epsfbox{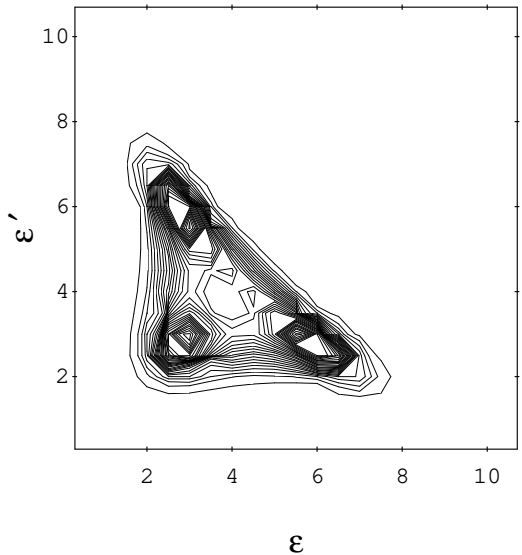}}}\fi
\caption{ Degeneracy of the three string system with 
a total energy $\frac{10}
{(\alpha^{\prime})^{1/2}}$. 
}
\label{degeneracy3}
\end{figure}

If the total energy is smaller than the mentioned value, the most probable distribution is
${\cal E}=\frac{E}{2}$ and then we have equipartition.  The strings will 
most probably be in the massless level or in the first Planck mass excitation 
because for these  two levels the degeneracy is not high enough to destabilize 
the system for energies lower than $E_0$. But when $E > E_0$ one of the  strings tends to absorbe
 more energy than the other, (see Fig.\ref{degeneracy2}).
 In this last regime equipartition is broken. 
Another way to observe this behaviour is by
looking at the density of states of the two string gas obtained through convolution, 
and comparing it with the single string one. As the energy grows both functions give 
the same termodynamics in the sense that for the same big total energy  both  
densities of states  produce the same thermal properties, see\footnote{In the plots, 
$\alpha'=1$ has been taken and the temperature has been scaled by $4\pi$ so
$T_H = 1$.} Fig.\ref{temp1}. 
In the next section we will find $T$ as a function of the energy.
The same behaviour can be seen for a three string gas, as in Fig.\ref{degeneracy3}. 
One could wonder  whether the fat (energetic) string would eventually capture
all the energy while the others disappear. To this purpose we can make a simple calculation
for the two-string example. We start from a concrete configuration of the gas with a total energy
$E$. We assume that one of the strings is energetic enough to approximate well its density of states
by the leading behaviour at high energies, and the other will be supposed to be in the massless
range. The degeneracy of this configuration would be
\begin{equation}
\label{sea}
\omega(E,{\cal E})\propto{\cal E}^{24}(E-{\cal E})^{-27/2} e^{\beta_H(E-{\cal E})}
\end{equation}
where ${\cal E}$ is  the energy of the light string.   
As a function of ${\cal E}$, $\omega(E,{\cal E})$ will increase if
\begin{equation}
\frac{24}{{\cal E}}-\beta_H+\frac{27}{2(E-{\cal E})}>0
\end{equation}
and decrease otherwise.  For a given value of 
${\cal E}$ if $\omega(E,{\cal E})$ grows (lessens) the most probable value of ${\cal E}$  would be
bigger (smaller) than the considered one. Note that if ${\cal E}<\frac{24}{\beta_H}$
then the condition 
will always be satisfied no matter how energetic the other string is. This means there will
always be a small string of at least that critical energy. This simple model allows us to induce
 that in a string gas there will exist, as the total energy becomes bigger, a very
energetic string and a low energetic string sea. In the next section we will make another 
explanation of this fact for a $n$-string gas using its thermal properties.

\section{Thermodynamics}
We start from some basic definitions of thermodynamical magnitudes in the microcanonical 
ensemble:
\begin{eqnarray}
T^{-1}(E) & = & \frac{\partial\log(\Omega(E))}{\partial E} = \frac{\Omega'(E)}{\Omega(E)} \nonumber \\
C_V^{-1}(E) & = & \frac{\partial T(E)}{\partial E}
\end{eqnarray}
The expressions above for the temperature clarify the problem of the infinite volume. 
Writing $\Omega(E)$ as a series in $n$ , the temperature becomes

\begin{equation}
T(E)=\lim_{V\rightarrow\infty}\frac{\sum_{n=1}^N {\cal A}(n,E) V^n}{\sum_{n=1}^N {\cal A}'(n,E) V^n}
= \frac{{\cal A}(N,E)}{{\cal A}'(N,E)}.
\vspace*{0.5cm}
\end{equation}
Then the only term one has to consider in order to calculate the thermal properties is the last one,
that is the term with the biggest $n=N$. In other words, we are
taking $V$ to infinity irrespective of the value of $E$.
 
In general the specific heat is a linear function of the number of subsystems
and so it is sometimes useful to define a specific heat per subsystem, as follows

\begin{equation}   
\tilde{C}_V^{-1}(u) = \frac{\partial T(u)}{\partial u}=\frac{N}{C_V(E)} 
\end{equation}

\begin{figure}
\let\picnaturalsize=N
\def\picsize{2in}
\def\picfilename{temp1.eps}
\ifx\nopictures Y\else{\ifx\epsfloaded Y\else\input epsf \fi
\let\epsfloaded=Y
\centerline{\ifx\picnaturalsize N\epsfxsize \picsize\fi
\epsfbox{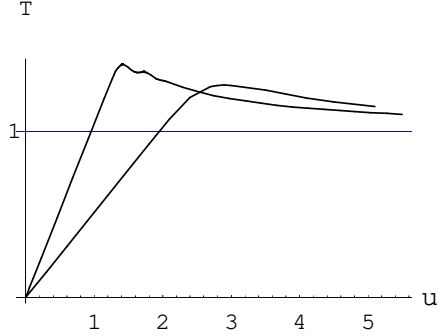}}}\fi
\caption{Temperature of the one and two-string gases with the same total energy. 
 Both systems have $E=2 u$, where $u$ is the energy per string in the two-string case. }
\label{temp1}
\end{figure}

\begin{figure}
%
%
\let\picnaturalsize=N
\def\picsize{2in}
\def\picfilename{temp2.eps}
\ifx\nopictures Y\else{\ifx\epsfloaded Y\else\input epsf \fi
\let\epsfloaded=Y
\centerline{\ifx\picnaturalsize N\epsfxsize \picsize\fi
\epsfbox{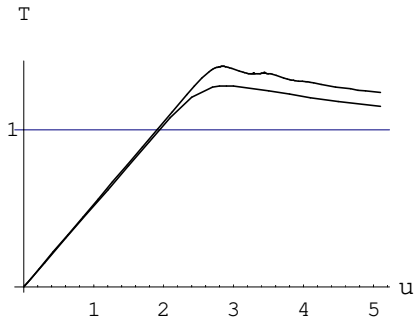}}}\fi

\vspace*{3.25cm}

\caption{Temperature of the single and two string gases with 
the same energy per string. }
%
%
\label{temp2}
\end{figure}

We have computed numerically the temperature for the single and the two-string gases. The results
are shown in Fig.\ref{temp1} and Fig.\ref{temp2}.

Some important features come up. First of all, one sees that 
both systems go from the low energy regime to the high energy one through a maximum. 
The explanation for the single string is clear; when the system is in the low energy regime 
everything happens like in a particle gas, that is, the specific heat is finite, positive and
 constant. However, when the amount of energy is enough for the string to reach the Planck mass 
level, this energy is in part used to create the mass and not to increase the kinetic energy so 
the temperature tends finally to decrease. In a system with only two levels the temperature 
would increase again because after creating the mass
the rest of the energy is employed in accelerating the particles, and so elevating the
temperature. In our system there is an infinite number of highly degenerated mass levels, each
one more than the previous, such that the energy will always be used to create ever more massive fields.
That is why the temperature gets lower.

 To calculate the high energy limit of the temperature, it is useful to approximate the density of states
 by the leading term in (\ref{highedensity}). It gives the well-known  expression for the
temperature:

\begin{equation}
T=\frac{E}{\beta_H E-\frac{27}{2}}
\end{equation}
which tends to $T_H$ when $E\rightarrow\infty$.

In the two-string gas the behaviour is practically the same except for the fact that the maximum
occurs at a lower temperature. This comes from the growth of the number of energy levels 
that are accesible
to each string when it borrows some energy from the other. The deviation from equipartition
is responsible for the cooling of the gas. The maximum appears approximately at a total energy 
$2 u_0$, that is, when each string has the critical energy assuming equipartition.
In both figures, we can see the two different regimes of the gas. When the energy is
large, the two-string gas behaves nearly as a single string with all the energy, just as the energy
distributions had predicted.

It is possible to generalize this high energy behaviour for a multiple string gas
 using the degeneracy of a configuration similar to the one in eq.(\ref{sea}). 
Let us suppose we have $N$ strings in the low energy regime and a fat one sharing 
a total energy $E$:

\begin{equation}
\omega(E,u)=\Omega_N(N u)\Omega_1(E-N u).
\end{equation}

Then taking the derivative with respect to $u$, which is now the density of energy per string,
it is easy to see that $\omega(E,u)$ will increase with $u$ only if $T_{fat}>T_{sea}$.
Since when the energy of the energetic string grows, the value of  $T_{fat}$ tends to
the Hagedorn temperature $T_H$, then in this situation the 
temperature of the other strings will also be $T_H$.
It seems that the string sea plays the r\^{o}le of a great heat bath in which the fat
string is contained.

We can put together all the information we have:  the single and two-string gases
density of states, the energy distribution of any number of strings and other physical hints
that allow us to present a pretty realistic model.

Let us imagine we have an infinite universe filled with a very rarefied and low energetic string gas.
Let us suppose we are able to introduce more and more energy at the same time that we observe the 
system evolution.
It would begin behaving like a particle gas with a positive specific heat per particle. As the
density of energy per string grows, the strings are able to occupy more and more massive states,
increasing the specific heat, as we have already explained for the two-string gas. At a certain
value of the energy, the temperature reaches the Hagedorn one. Until that point, equipartition
applies, but if we tried to increase slightly the energy of the gas, the probable distribution would
change in such a way that a single string would come out from the gas and absorb all
the energy added. In other words, the most probable number of strings absorbing energy is one.

It is very important to note that since we are considering an infinite volume universe, the number
of strings and the total energy of the system are also infinite. Hence, it is only possible to deal
with densities per unit volume or per subsystem. Both densities are simply related at least in the 
particle
case, as we have seen before. For energies higher than the critical one, the density of strings is
always the same as no energy is absorbed or emitted by the string sea.

\begin{figure}
\let\picnaturalsize=N
\def\picsize{2in}
\def\picfilename{tentati.eps}
\ifx\nopictures Y\else{\ifx\epsfloaded Y\else\input epsf \fi
\let\epsfloaded=Y
\centerline{\ifx\picnaturalsize N\epsfxsize \picsize\fi
\epsfbox{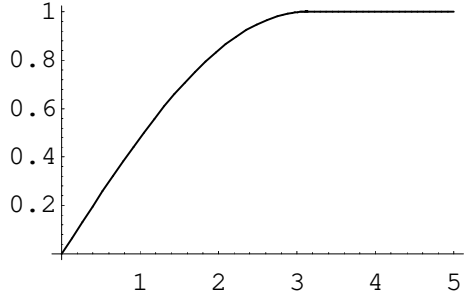}}}\fi
\caption{Tentative plot of temperature versus energy per string }
\label{tenta}
\end{figure}

It is clear that the fat string appears with an already infinite amount of
energy, and it can only exist being infinitely energetic. Therefore its temperature will always be
Hagedorn. We find no negative specific heat for any value of the energy. We do, however, find a divergent
specific heat for energies per string bigger than a certain critical point (approximately $u_0$). This is not a
surprise because when a finite maximum temperature of a system is expected, the specific heat diverges. For
example, the only difference with the open string case is that this divergence in $C_V$
 appears at a finite value
of $u$. 
One could wonder whether this divergence appears abruptly or in a continuous way. We have
seen that when one makes one convolution, the maximum temperature gets lower. The corresponding maxima 
are smooth and no discontinuity in $T$ is observed. That is why we think that the natural 
extrapolation for a large number of convolutions is that the system will keep this behaviour,
and the maximum, that we already know is Hagedorn, will be reached smoothly. We present a
tentative plot of the qualitative thermal behaviour of the system in Fig.\ref{tenta}.

\section{Conclusions}
In this paper we have revisited the thermal behaviour of a bosonic string gas in the microcanonical
ensemble. We have introduced some mathematical tools which seem to be useful to make the inverse Laplace
transorm. The r\^{o}le of the analytical parts have manifested as important to grant the convergence of
the series which allowed us to account for the quantum statistics
effetcs in the single string density of states at high energies. However, the physical range of
application of (\ref{highedensity}) is not known because it does not contain the
infrared part of the free energy associated in the $\beta$ complex plane with singularities on the
imaginary axis.
On the other hand this expression is completely useless in order to calculate the multiple string
density. This is due to the need of the convolution theorem, that we see as the most physical 
way to compute the multiple string case.
This enforces us to try to compute a valid expression for the density of states for the whole energy
range. We succeeded making that for the single string and, by application of the convolution theorem,
we observe a behaviour from which we induce that for infinite volume, summing up over an infinite
number of subsystems, no negative specific heat phase appears. On the contrary, at high energies, we
have a system which absorbs any amount of energy without increasing its temperature, so the
limiting temperature $T_H$,
 is actually a maximum one for the universe. What seems to be clear to
us is that the description of this phase as one with, not only a fat string, but
also with a sea of a large number of low-energy strings is not adequate to  the degrees of
freedom we have at hand. It is only the result of trying to enforce a 
description of these degrees
of freedom in terms of strings 
when no equilibrium is posible among them as subsystems. To be more concrete, we do not believe an alternative description
of these degrees of freedom should change the fact that, perturbatively at least, Hagedorn is the 
maximum temperature of the system. 

The study we have presented  can be easily extended to Superstrings.
The next refinement of the thermodynamics 
would be considering non-perturbative effects as D-Instantons. A calculation 
of the influence of these effects on the one loop free energy has been done in \cite{BarboVazquez},
where the authors  have also obtained an expression for the high energy density of states for
this case.
This type of calculation is along the lines of the works in \cite{DeoAxeMac}
in which no infinite volume limit is actually taken, but what they do amounts to
taking a high energy  limit for each $\Omega_n$ in (\ref{omegan}) then substituting the energy dependence for that of
the single string in such a way that finally they approximate $\Omega(E)$ by the asymptotic expression for
the single string density times
the exponential of $V$. So  no limit in $V$ is in fact needed. 
In this way no phase transition can then be captured because no critical
limit is taken. Obviously this procedure is completely different from ours
ab initio.

We believe that to include D-Instantons effects in our picture, it is necessary   
to compute once more the complete density of states for the single string with this corrections, 
and then, by convolution, obtain the multiple 
string gas.  

\section*{Acknowledgements}
We are greatful to J. L. F. Barb\'{o}n and M.A. V\'{a}zquez Mozo for discussions. Thanks to
 E. \'{A}lvarez for reading the manuscript and to J.M.
Noriega Antu\~{n}a for helping us to find an indispensable book. Jes\'{u}s and Marco are in debt with
their parents for the economic support.

\end{document}